\documentstyle[12pt,epsf]{article}

\def\npb#1#2#3{    {\it Nucl. Phys. }{\bf B #1} (19#2) #3}
\def\plb#1#2#3{    {\it Phys. Lett. }{\bf B #1} (19#2) #3}
\def\prd#1#2#3{    {\it Phys. Rev. }{\bf D #1} (19#2) #3}
\def\prep#1#2#3{   {\it Phys. Rep. }{\bf #1} (19#2) #3}
\def\prl#1#2#3{    {\it Phys. Rev. Lett. }{\bf #1} (19#2) #3}

\def\zpc#1#2#3{    {\it Zeit. f\"ur Physik }{\bf C #1} (19#2) #3}

\newcommand{\beqn}{\begin{eqnarray}}
\newcommand{\beq}{\begin{equation}}
\newcommand{\eeqn}{\end{eqnarray}}
\newcommand{\eeq}{\end{equation}}
\newcommand{\nn}{\nonumber}

\begin{document}
\begin{titlepage}
\begin{flushright}
{
ROM2F/96/48 \\
September 1996
}
\end{flushright}
\vskip 1cm
\centerline {\Large{\bf The Beauty of SUSY$^{*}$}}
\normalsize
 
\vskip 1cm
\centerline {A. Masiero}
\centerline {\it Dip. di Fisica, Univ. di Perugia and}
\centerline {\it INFN, sez. di Perugia, Via Pascoli, I-06100 Perugia, Italy}
\vskip 0.5cm
\centerline{\it and}
\vskip 0.5cm
\centerline {L. Silvestrini}
\centerline {\it Dip. di Fisica, Univ. di Roma ``Tor Vergata" and}
\centerline {\it INFN, sez. di Roma II,}
\centerline {\it Via della Ricerca Scientifica, 1, I-00133 Roma, Italy}

\vskip 1.5cm
 
\begin{abstract}
$B$ physics represents a privileged place to look for supersymmetry (SUSY) 
through its virtual effects. Here we discuss rare $B$ decays ($b \to s 
\gamma$, $b \to s g$, $b \to s l^{+} l^{-}$) and $B-\bar{B}$ oscillations 
in the context of low-energy SUSY. We outline the variety of predictions that 
arise according to the choice of the SUSY extension ranging from what we call 
the ``minimal" version of the MSSM to models without flavour universality or 
with broken R-parity. In particular, we provide a model-independent 
parameterization of the SUSY FCNC effects which is useful in tackling the 
problem in generic low-energy SUSY. We show how rare $B$ physics may be 
complementary to direct SUSY searches at colliders, in particular for what 
concerns extensions of the most restrictive version of the MSSM.
\end{abstract}
\vskip 1cm
\centerline{$^{*}$ Talk given by A. Masiero at BEAUTY 96, Rome,
Italy, 17-21 June 1996.}
\vfill
\end{titlepage}

\newpage
\section{Introduction}
Flavour Changing Neutral Current (FCNC) phenomena represent a major test for 
any extension of the Standard Model (SM) which is characterized by an energy 
scale $\Lambda$ close to the electroweak scale. Low-energy supersymmetry 
(SUSY) is no exception in this way: its prediction of new particles carrying 
flavour numbers with masses not exceeding a few TeV's (i.e., $\Lambda$ $<$ few 
TeV's in this case) makes the indirect search of SUSY manifestations through 
virtual effects in FCNC processes of utmost interest.

The potentiality of probing SUSY in FCNC phenomena was readily realized when 
the era of SUSY  phenomenology started in the early 80's \cite{susy2}.
 In particular, the 
major implication that the scalar partners of quarks of the same electric 
charge but belonging to different generations had to share a remarkable high 
mass degeneracy was emphasized. 

Throughout the large amount of work  in this last decade it became clearer 
and clearer that generically talking of the implications of low-energy SUSY on 
FCNC may be rather misleading. We have a minimal SUSY extension of the SM, the 
so-called Minimal Supersymmetric Standard Model (MSSM) \cite{susy1}, 
where the FCNC 
contributions can be computed in terms of a very limited set of unknown 
new SUSY parameters. Remarkably enough, this minimal model succeeds to pass all 
the set of FCNC tests unscathed. To be sure, it is possible to severely 
constrain the SUSY parameter space, for instance using $b 
\to s \gamma$ in a way which is complementary to what is achieved by direct 
SUSY searches at colliders.

However,  the MSSM is by no means equivalent to low-energy SUSY. First, there 
exists an interesting large class of SUSY realizations where the customary 
discrete R-parity (which is invoked to suppress proton decay) is replaced by 
other discrete symmetries which allow either baryon or lepton violating terms 
in the superpotential. But, even sticking to the more orthodox view of 
imposing R-parity, we are still left with a large variety of extensions of the 
MSSM at low energy. The point is that low-energy SUSY ``feels" the new physics 
at the superlarge scale at which supergravity  (i.e., local supersymmetry) 
broke down. In this last couple of years we have witnessed an increasing 
interest in supergravity realizations without the so-called flavour 
universality of the terms which break SUSY explicitly. Another class of 
low-energy SUSY realizations which differ from the MSSM in the FCNC sector 
is obtained from SUSY-GUT's. The interactions involving superheavy particles 
in the energy range between the GUT and the Planck scale bear important 
implications for the amount and kind of FCNC that we expect at low energy.

After an initial effort on the study of  FCNC SUSY effects in kaon physics it 
became clear that $B$ physics represents the new (and, for many aspects, more 
promising) frontier for probing SUSY through FCNC effects in the hadronic 
sector. There has already been an intense research activity 
in the realm of rare FCNC $B$ decays and SUSY. The simultaneous progress on the 
experimental side and, even more, the prospects that new $B$ facilities open up 
in these coming years make these studies of enormous interest in our effort 
to detail the structure (and the existence!) of low-energy SUSY.

In this talk I will review some of the most recent work along these lines, in 
particular distinguishing  the situation concerning the MSSM and other 
low-energy SUSY realizations.

\section{FCNC in SUSY without R-Parity}

It is well known that in the SM case the imposition of gauge symmetry and the 
usual gauge assignment of the 15 elementary fermions of each family lead to 
the automatic conservation of baryon (B) and lepton (L) numbers 
(this is true at any order in 
perturbation theory).

On the contrary, imposing in addition to the usual $SU(3)\otimes SU(2) \otimes 
U(1)$ gauge symmetry an N=1 global SUSY does not prevent the appearance of 
terms which explicitly break B or L \cite{weinb}.
 Indeed, the superpotential reads:
\beqn
W&=&h^U Q H_{U}u^c + h^D Q H_{D} d^c + h^L L H_D e^c + \mu H_U H_D \nn \\
&+& \mu^\prime H_{U} L + \lambda^{\prime \prime}_{ijk}u^c_{i}d^c_{j}d_{k}^c +
\lambda^{\prime}_{ijk}Q_{i}L_{j}d_{k}^c + \lambda_{ijk}L_{i}L_{j}e_{k}^c \, ,
\label{superp}
\eeqn
where the chiral matter superfields $Q$, $u^c$, $d^c$, $L$, $e^c$, $H_{U}$ and 
$H_{D}$ transform under the above gauge symmetry as:
\beqn
&\,&Q\equiv (3,2,1/6); \qquad u^c\equiv (\bar{3},1,-2/3);\qquad d^c\equiv
(\bar{3},1,1/3);\\
&\,& L\equiv (1,2,-1/2); \; \; e^c \equiv (1,1,1); \;\; H_{U}\equiv 
(1,2,1/2); \;\; H_{D}\equiv (1,2,-1/2). \nn
\label{qnumbers}
\eeqn
The couplings $h^U$, $h^D$, $h^L$ are $3\times 3$ matrices in the generation 
space; $i$, $j$ and $k$ are generation indices. Using the product of 
$\lambda^\prime$ and $\lambda^{\prime \prime}$ couplings it is immediate to 
construct four-fermion operators leading to proton decay through the exchange 
of a squark. Even if one allows for the existence of $\lambda^\prime$ and 
$\lambda^{\prime \prime}$ couplings only involving the heaviest generation, 
one can show that the bound on the product $\lambda^\prime \times 
\lambda^{\prime \prime}$ of these couplings is very severe (of $O(10^{-7})$)
\cite{smirnov}.

A solution is that there exists a discrete symmetry, B-parity \cite{b}, 
which forbids 
the B violating terms in eq.~(\ref{superp}) which are proportional to 
$\lambda^{\prime \prime}$. In that case it is still possible to produce 
sizeable effects in FC $B$ decays. For instance using the product of 
$\lambda^\prime_{3jk}\lambda_{ljl^c}$ one can obtain $b \to s \,(d) + l l^c$ 
taking $k=2 \, (1)$ and through the mediation of the sneutrino of the $j$-th 
generation. Two general features of these R-parity violating contributions 
are:
\begin{enumerate}
\item we completely lose any correlation to the CKM elements. For instance, in 
the above example, the couplings $\lambda^\prime$ and $\lambda$ have nothing 
to do with the usual angles $V_{tb}$ and $V_{ts}$ which appear in $b \to s l^+ 
l^-$ in the SM;
\item we also lose correlation among different FCNC processes which are 
tightly correlated in the SM. For instance, in our example $b \to d l^+ l^-$ 
would depend on $\lambda^\prime$ and $\lambda$ parameters which are different 
from those appearing in $B_{d}-\bar{B}_{d}$ mixing.
\end{enumerate}

In this context it is difficult to make predictions given the arbitrariness of 
the large number of $\lambda$ and $\lambda^\prime$ parameters. There exist 
bounds on each individual coupling (i.e. assuming all the other L violating 
couplings are zero) \cite{barger}.
With some exception, they are not very stringent for the third generation 
(generally of $O(10^{-1})$), 
hence allowing for conspicuous effects. Indeed, one 
may think of using the experimental bounds on rare $B$ decays to put severe 
bounds on products of L violating couplings.

Obviously, the most practical way of avoiding any threat of B and L violating 
operators is to forbid \underline{all} such terms in eq.~(\ref{superp}). This 
is achieved by imposing the usual R matter parity. This quantum number reads 
$+1$ over every ordinary particle and $-1$ over SUSY partners. We now turn to 
rare $B$ decays in the framework of low energy SUSY with R-parity.

\section{Model-independent analysis of FCNC processes in SUSY}

Given a specific SUSY model it is in principle possible to make a full 
computation of all the FCNC phenomena in that context. However, given the 
variety of options for low-energy SUSY which was mentioned in the Introduction 
(even confining ourselves here to models with R matter parity), it is 
important to have a way to extract from the whole host of FCNC processes a set 
of upper limits on quantities which can be readily computed in any chosen SUSY 
frame.

The best model-independent parameterization of FCNC effects is the so-called 
mass insertion approximation \cite{mins}. 
It concerns the most peculiar source of FCNC SUSY contributions that do not 
arise from the mere supersymmetrization of the FCNC in the SM. They originate 
from the FC couplings of gluinos and neutralinos to fermions and 
sfermions~\cite{FCNC}. One chooses a basis
for the fermion and sfermion states where all the couplings of these particles
to neutral gauginos are flavour diagonal, while the FC is exhibited by the
non-diagonality of the sfermion propagators. Denoting by $\Delta$ the
off-diagonal terms in the sfermion mass matrices (i.e. the mass terms relating
sfermion of the same electric charge, but different flavour), the sfermion
propagators can be expanded as a series in terms of $\delta = \Delta/
\tilde{m}^2$
where $\tilde{m}$ is the average sfermion mass. 
As long as $\Delta$ is significantly smaller than $\tilde{m}^2$, 
we can just take
the first term of this expansion and, then, the experimental information
concerning FCNC and CP violating phenomena translates into upper bounds on 
these $\delta$'s \cite{deltas}.
 
  Obviously the above mass insertion method presents the major advantage that
 one does not need the full diagonalization of the sfermion mass matrices to
 perform a test of the SUSY model under consideration in the FCNC sector. It is
 enough to compute ratios of the off-diagonal over the diagonal entries of the
 sfermion mass matrices and compare the results with the general bounds on the
 $\delta$'s that we provide here from all available experimental information.

  There exist four different 
$\Delta$ mass insertions connecting flavours $i$ and $j$
 along a sfermion propagator: $\left(\Delta_{ij}\right)_{LL}$, 
$\left(\Delta_{ij}\right)_{RR}$, $\left(\Delta_{ij}\right)_{LR}$ and 
$\left(\Delta_{ij}\right)_{RL}$. The indices $L$ and $R$ refer to the 
helicity of 
the
 fermion partners. The size of these $\Delta$'s can be quite different. For
 instance, it is well known that in the MSSM case, only the $LL$ mass insertion
 can change flavour, while all the other three above mass insertions are flavour
 conserving, i.e. they have $i=j$. In this case to realize a $LR$ or $RL$ 
flavour
 change one needs a double mass 
insertion with the flavour changed solely in a $LL$
 mass insertion and a subsequent flavour-conserving $LR$ mass insertion. 
Even worse
 is the case of a FC $RR$ transition: in the MSSM this can be accomplished only
 through a laborious set of three mass insertions, two flavour-conserving $LR$
transitions and an $LL$ FC insertion. 
  Instead of the dimensional quantities $\Delta$ it is more
useful to provide bounds making use of dimensionless quantities, $\delta$, 
that are obtained dividing the mass insertions by an average sfermion mass.

The FCNC processes in $B$ physics which provide the best bounds on the 
$\delta_{23}$ and $\delta_{13}$ FC insertions are $b \to s \gamma$ and $B_{d}-
\bar{B}_{d}$, respectively.

\begin{figure}
    \begin{center}
    \epsfysize=12truecm
    \leavevmode\epsffile{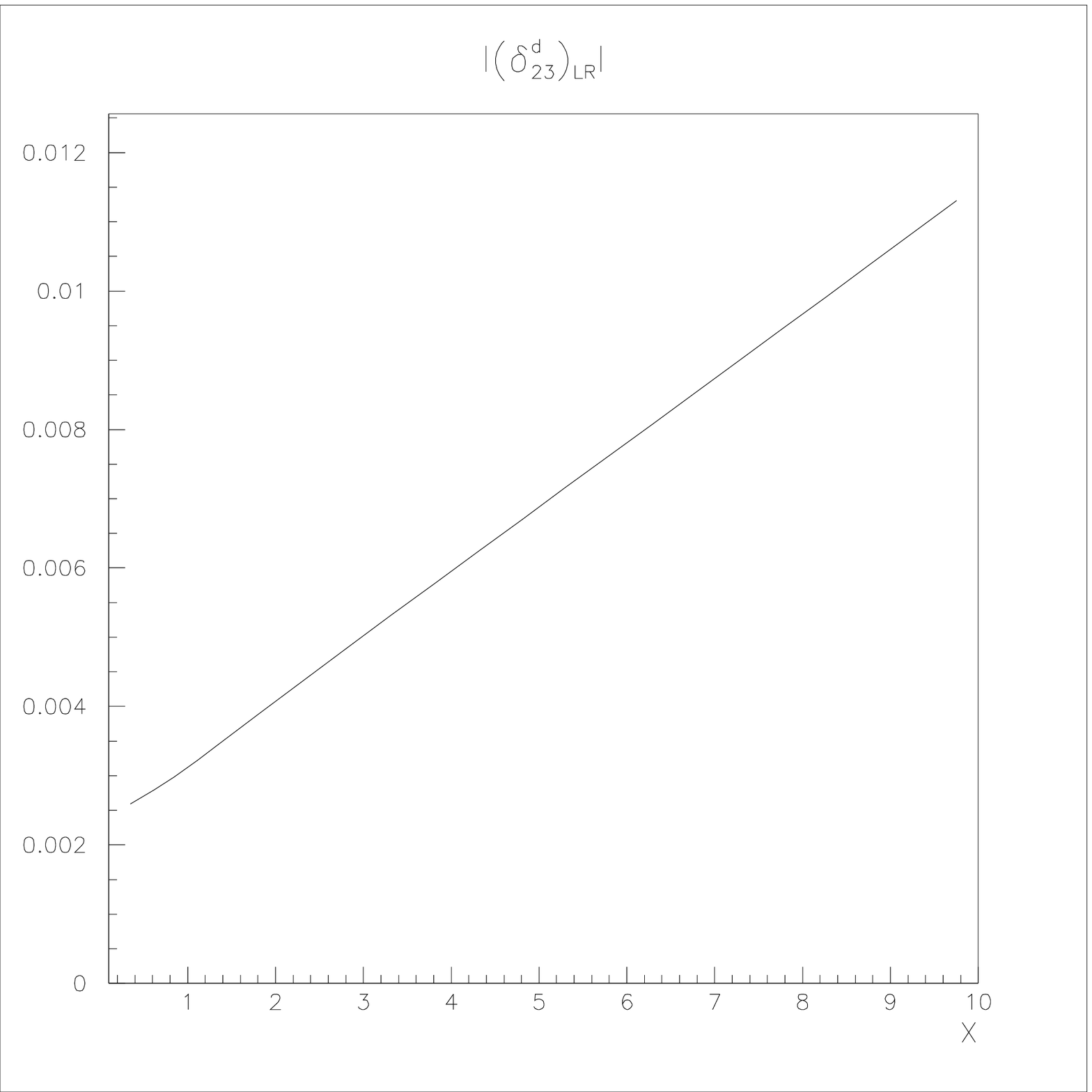}
    \end{center}
    \caption[]{The $\left\vert\left(\delta^d_{23}\right)_{LR}\right\vert$ 
     as a function
     of $x=m_{\tilde{g}}^2/m_{\tilde{q}}^2$, for  an average squark mass 
     $m_{\tilde{q}}=100\mbox{GeV}$. For different values of $m_{\tilde{q}}$, 
     the limits scale as $m_{\tilde{q}}(\mbox{GeV})/100$.}
\label{bsglr}
\end{figure}

The process $b \to s \gamma$ requires a helicity flip. In the presence of a 
$\left(\delta^d_{23}\right)_{LR}$ mass insertion we can realize this flip in 
the gluino running in the loop. On the contrary, the $\left(
\delta^d_{23}\right)_{LL}$ insertion requires the helicity flip to occur in 
the external $b$-quark line. Hence we expect a stronger bound on the 
$\left(\delta^d_{23}\right)_{LR}$ quantity. Indeed, this is what happens: 
$\left(\delta^d_{23}\right)_{LL}$ is essentially not bounded, while 
$\left(\delta^d_{23}\right)_{LR}$ is limited to be $<10^{-3}-10^{-2}$ 
according to the average squark and gluino masses (see fig.~\ref{bsglr}). 
Given the upper 
bound on $\left(\delta^d_{23}\right)_{LR}$ from $b \to s \gamma$, it turns out 
that the quantity $x_{s}$ of the $B_{s}-\bar{B}_{s}$ mixing receives 
contributions from this kind of mass insertions which are very tiny. The only 
chance to obtain large values of $x_s$ is if $\left(\delta^d_{23}\right)_{LL}$ 
is large, say of $O(1)$. In that case $x_s$ can easily jump up to values of $O
(10^{2})$ or even larger.

As for the mixing $B_{d}-\bar{B}_{d}$, we obtain 
\beqn
\sqrt{\left\vert {\mbox Re} \left(\delta^d_{13}\right)^{2}_{LL}\right\vert}
&<&4.6 \cdot 10^{-2}\, ; \nn \\
\sqrt{\left\vert {\mbox Re} \left(\delta^d_{13}\right)^{2}_{LR}\right\vert}
&<&5.6 \cdot 10^{-2}\, ; \nn \\
\sqrt{\left\vert {\mbox Re} \left(\delta^d_{13}\right)_{LL}\left(
\delta^d_{13}\right)_{RR}\right\vert}
&<&1.6 \cdot 10^{-2}\, ; 
\label{limbdbdb}
\eeqn
for $x\equiv m^{2}_{\tilde{g}}/m^{2}_{\tilde{q}}=0.3$ with $m_{\tilde{q}}=500$ 
 GeV. The above bounds scale with $m_{\tilde{q}}$(GeV)$/500$ for different 
values of $m_{\tilde{q}}$ (at fixed $x$).\\
Then, imposing the bounds~(\ref{limbdbdb}), we can obtain the largest possible 
value for BR($b \to d \gamma$) through gluino exchange. As expected, the 
$\left( \delta^{d}_{13}\right)_{LL}$ insertion leads to very small values of 
this BR of $O(10^{-7})$ or so, whilst the $\left( \delta^{d}_{13}\right)_{LR}$ 
insertion allows for BR($b \to d \gamma$) ranging from few times $10^{-4}$ up 
to few times $10^{-3}$ for decreasing values of $x=m^{2}_{\tilde{g}}/
m^{2}_{\tilde{q}}$. As reminded by Ali at this 
meeting, in the SM we expect BR($b \to d \gamma$) to be typically $10-20$ 
times smaller than BR($b \to s \gamma$), i.e. BR($b \to d \gamma)=(1.7\pm 0.85
)\times 10^{-5}$. Hence a large enhancement in the SUSY case is conceivable if 
$\left( \delta^{d}_{13}\right)_{LR}$ is in the $10^{-2}$ range. Notice that in 
the MSSM we expect $\left( \delta^{d}_{13}\right)_{LR}<m^{2}_{b}/
m^{2}_{\tilde{q}}\times V_{td}<10^{-6}$, hence with no hope at all of a 
sizeable contribution to $b \to d \gamma$.

However, as we shall see in Sect.~\ref{sec:rare}, sizeable deviations from the 
expected values of the $\delta$ quantities in the MSSM are possible in SUSY 
schemes which are obtained as the low-energy limit of $N=1$ supergravities 
with a GUT structure and/or non-universal soft breaking terms.

\section{Rare $B$ decays in the MSSM and beyond}
\label{sec:rare}

Although the name seems to indicate a well-defined particle model, actually 
MSSM denotes at least two quite different classes of low-energy SUSY models. 
In its most restrictive meaning it denotes the minimal SUSY extension of the 
SM (i.e. with the smallest needed number of superfields) with R-parity, 
radiative breaking of electroweak symmetry, universality of the soft breaking 
terms and simplifying relations at the GUT scale among SUSY parameters. In 
this ``minimal" version the MSSM exhibits only four free parameters in 
addition to those of the SM. Moreover, some authors impose specific relations 
between the two parameters $A$ and $B$ that appear in the trilinear and 
bilinear scalar terms of the soft breaking sector further reducing the number 
of SUSY free parameters to three. Then, all SUSY masses are just function of 
these few independent parameters and, hence, many relations among them exist. 
Obviously this very minimal version of the MSSM can be very predictive. The 
most powerful constraint on this minimal model in the FCNC context comes from 
$b \to s \gamma$.

In SUSY there are five classes of one-loop diagrams which contribute 
to FCNC $B$ 
processes. They are distinguished according to the virtual particles running 
in the loop: W and up-quarks, charged Higgs and up-quarks, charginos and 
up-squarks, neutralinos and down-squarks, gluinos and down-squarks. It turns 
out that, at least in this ``minimal" version of the MSSM, the charged Higgs 
and chargino exchanges yield the dominant SUSY contributions. As for $b \to s 
\gamma$ the situation can be summarized as follows. The CLEO measurement 
yields BR$(B \to X_{s}\gamma)=(2.32 \pm 0.67)\times 10^{-4}$ \cite{cleo}. 
On the 
theoretical side we are going to witness a major breakthrough 
with the 
computation of the next-to-leading logarithmic result for the BR. This is
achieved thanks to the calculation of the $O(\alpha_{s})$ matrix elements 
\cite{greub} and 
of the next-to-leading order Wilson coefficients at $\mu \simeq m_{b}$ 
\cite{misiak}. The 
result quoted by Greub and Hurth \cite{hurth} 
is BR$(B \to X_{s} \gamma)=(3.25 \pm 0.50) 
\times 10^{-4}$ in the SM with $m_{t}=(170 \pm 15)$ GeV and $m_{b}/2 \le \mu
\le 2 m_{b}$. A substantial improvement also on the experimental error is 
foreseen for the near future. Hence $b \to s \gamma$ is going to constitute 
the most relevant place in FCNC $B$ physics to constrain SUSY at least before 
the advent of $B$ factories. So far this process has helped in ruling out 
regions of the SUSY parameter space which are even larger than those excluded 
by LEP I and it is certainly going to be complementary to what LEP II is 
expected to do in probing the SUSY parameter space. After the detailed 
analysis in 1991 \cite{bertol} 
for small values of $\tan \beta$, there have been recent 
analyses \cite{barb}
  covering the entire range of $\tan \beta$ and including also other 
technical improvements (for instance radiative corrections in the Higgs 
potential). It has been shown \cite{vissani} 
that the exclusion plots are very sensitive also 
to the relation one chooses between A and B. It should be kept in mind that 
the ``traditional" relation $B=A-1$ holds true only in some simplified version 
of the MSSM. A full discussion is beyond the scope of this talk and so we 
refer
the interested readers to the vast literature which exists on the subject. 

The constraint on the SUSY parameter space of the ``minimal" version of the 
MSSM greatly affects also the potential departures of this model from the SM 
expectation for $b \to s l^+ l^-$. The present limits on the exclusive 
channels BR$(B^{0} \to K^{*0} e^{+} e^{-})_{CLEO}<1.6 \times 10^{-5}$ 
\cite{cleo2} and BR$(
B^{0} \to K^{*0} \mu^{+} \mu^{-})_{CDF}<2.1 \times 10^{-5}$ \cite{cdf} 
are within an 
order of magnitude of the SM predictions. On the theoretical side, it has 
been estimated that the evaluation of $\Gamma (B\to X_{s}l^{+} l^{-})$ in the 
SM is going to be affected by an error which cannot be reduced to less than 
$10-20 \%$ due to uncertainties in quark masses and interference effects from 
excited charmonium states \cite{ligeti}. 
It turns out that, keeping into account the bound 
on $b \to s \gamma$, in the MSSM with universal soft breaking terms a $20 \%$ 
departure from the SM expected BR is kind of largest possible value one can 
obtain \cite{cho}. 
Hence the chances to observe a meaningful deviation in this case are 
quite slim. However, it has been stressed that in view of the fact that three 
Wilson 
coefficients play a relevant role in the effective low-energy Hamiltonian 
involved in $b \to s \gamma$ and $b \to s l^{+} l^{-}$, a third observable in 
addition to BR$(b \to s \gamma)$ and BR$(b \to s l^{+}l^{-})$ is needed.
This has been identified in some asymmetry of the emitted leptons (see
 refs.~\cite{cho, ali} for two different choices of such asymmetry). 
This quantity, even 
in the ``minimal" MSSM, may undergo a conspicuous deviation from its SM 
expectation and, hence, hopes of some manifestation of SUSY, even in this 
minimal realization, in $b \to s l^{+} l^{-}$ are still present.

Finally, also for the $B_{d}-\bar{B}_{d}$ mixing, in the above-mentioned 
analysis of rare $B$ physics in the MSSM with universal soft breaking terms 
\cite{bertol} 
it was emphasized that, at least in the low $\tan \beta$ regime, one cannot 
expect an enhancement larger than $20\%-30\%$ over the SM prediction 
(see also ref.~\cite{kurimoto}). Moreover 
it was shown that $x_{s}/x_{d}$ is expected to be the same as in the SM.

It should be kept in mind that the above stringent results strictly depend not 
only on the minimality of the model in terms of the superfields that are 
introduced, but also on the ``boundary" conditions that are chosen.
All the low-energy SUSY masses are computed in terms of the $M_{Pl}$ four 
SUSY parameters through the RGE evolution. If one relaxes this tight 
constraint on the relation of the low-energy quantities and treats the masses 
of the SUSY particles as independent parameters, then much more freedom is 
gained. This holds true even if flavour universality is enforced. For 
instance,
 BR$(b \to s \gamma
)$ and $\Delta m_{B_{d}}$ may vary a lot from the SM expectation, in 
particular in regions of moderate SUSY masses \cite{brignole}.

Moreover, flavour universality is by no means a prediction of low-energy SUSY. 
The absence of flavour universality of soft-breaking terms may result from 
radiative effects at the GUT scale or from effective supergravities derived 
from string theory. For instance, FCNC contributions in a minimal SUSY SO(10) 
model might be comparable to some of the upper bounds on the FCNC $\delta$ 
quantities given above \cite{strumia}. In the non-universal case, 
 BR$(b \to s l^{+} l^{-})$ 
is strongly affected by this larger freedom in the parameter space. There are 
points of this parameter space where the nonresonant BR$(B \to X_{s} e^{+} 
e^{-})$ and BR$(B \to X_{s} \mu^{+}\mu^{-})$  are enhanced by up to $90 \%$ and 
$110 \%$ while still respecting the constraint coming from $b \to s \gamma$
\cite{cho}.

Finally, let us add a short comment on another rare $B$ decay, $b \to s g$, 
which has attracted some attention in these last years for its potentiality to 
increase the $b$ hadronic width, hence lowering the $b$ semileptonic branching 
ratio. It was recently noticed in refs.~\cite{kagan, ciu} that a 
BR$(b \to s g)$ close 
to $10 \%$ could simultaneously solve the two famous problems in $B$ physics 
of the semileptonic branching ratio and of the number of charms per $B$ decay 
(charm counting). At first sight, one could object that it is unlikely to have 
such a large BR$(b \to s g)$ given that its close ``friend" BR$(b \to s 
\gamma)$ is at the $10^{-4}$ level. However it was shown in ref.~\cite{ciu} 
that there exists an admittedly small region of the SUSY parameter space 
where, indeed, BR$(b \to s g)$ is as high as $10 \%$ without conflicting with 
the measured value of BR$(b \to s \gamma)$. However, at this meeting S. Stone 
and some of his collaborators have emphasized that a BR$(b \to s g)$ at the 
$10 \%$ level is in difficulty with the results of searches for the 
$\Phi$ from $B \to X_{s} \Phi$.

\section{Conclusions}

We summarize the results reported in this talk in the following three points.
\begin{enumerate}
\item First, a warning: under the commonly used expression of low-energy SUSY 
there exists actually a large choice of models (with or without R parity, with 
or without flavour universality of the soft breaking terms, with or without 
GUT assumptions,\dots) which lead to quite different 
implications for rare $B$ decays.
\item What we called here the ``minimal" version of the MSSM, i.e. its most 
restrictive version with only four independent parameters, is mainly 
constrained by $b \to s \gamma$, with little hope of significant departures 
from SM in other FCNC $b$ physics (however, some exception is possible, like, 
for instance, lepton asymmetries in $b \to s l^{+} l^{-}$).
\item Extensions of the above ``minimal" version of the MSSM with 
non-universality, or with SUSY-GUT's, 
have room for conspicuous departures from the SM in $b \to s l^{+} 
l^{-}$ and $b \to s g$.
\end{enumerate}

The hope is that the advent of $B$ factories may promote $B$ physics to a 
ground for ``precision tests" of new physics analogously to what has been done 
for the LEP Z factory.

\section*{Acknowledgements}

We thank F. Gabbiani and E. Gabrielli who collaborated with us in the 
model-independent 
analyses that we present here. Helpful discussions with A. Ali and 
G.F. Giudice are kindly acknowledged. Finally, we wish to thank the organizers 
of BEAUTY '96 for giving us the opportunity of discussing $B$ physics in a 
stimulating and pleasant atmosphere.

\newpage

\begin{thebibliography}{99}

\bibitem{susy2}
J. Ellis and D.V. Nanopoulos, \plb{110}{82}{44};\\
R. Barbieri and R. Gatto, \plb{110}{82}{211}.

\bibitem{susy1} 
For a phenomenologically oriented review, see:\\
P. Fayet and S. Ferrara, \prep{32C}{77}{249};\\
H.P. Nilles, \prep{110C}{84}{1}.\\
H.E. Haber and G.L Kane, \prep{117C}{87}{1};\\
For spontaneously broken N=1 supergravity, see:\\
E. Cremmer, S. Ferrara, L. Girardello and A. Van Proeyen, \npb{212}{83}{413}
and references therein.\\
P. Nath, R. Arnowitt and A.H. Chamseddine, {\it Applied N=1 Supergravity} (
World Scientific, Singapore, 1984);\\
A.G. Lahanas and D.V. Nanopoulos, \prep{145C}{87}{1}.

\bibitem{weinb}
S. Weinberg, \prd{26}{82}{287};\\
N. Sakai and T. Yanagida, \npb{197}{82}{533}.

\bibitem{smirnov}
A.Y. Smirnov and F. Vissani, \plb{380}{96}{317}.

\bibitem{b}
C. Aulakh and R.N. Mohapatra, \plb{119}{83}{136};\\
L.J. Hall and M. Suzuki, \npb{231}{84}{419};\\
I.H. Lee, \npb{246}{84}{120};\\
J. Ellis et al., \plb{150}{85}{142}.\\
For a discussion in the string context, see:\\
L.E. Ibanez and G.G. Ross, \npb{368}{92}{3}.

\bibitem{barger}
V. Barger, G.F. Giudice and T. Han, \prd{40}{89}{2987};\\
K. Enqvist, A. Masiero and A. Riotto, \npb{373}{92}{95}.

\bibitem{mins}
L.J. Hall, V.A. Kostelecky and S. Raby, \npb{267}{86}{415}.

\bibitem{FCNC}
M.J. Duncan, \npb{221}{83}{285};\\
J.F. Donoghue, H.P. Nilles and D. Wyler, \plb{128}{83}{55};\\
A. Bouquet, J. Kaplan and C.A. Savoy, \plb{148}{84}{69}.

\bibitem{deltas}
F. Gabbiani and A. Masiero, \npb{322}{89}{235};\\
J.S. Hagelin, S. Kelley and T. Tanaka, \npb{415}{94}{293};\\
E. Gabrielli, A. Masiero and L. Silvestrini, \plb{374}{96}{80};\\
F. Gabbiani, E. Gabrielli, A. Masiero and L. Silvestrini, \npb{477}{96}{321}.

\bibitem{cleo}
M.S. Alam et al. (CLEO collab.), \prl{74}{95}{2885}.

\bibitem{greub}
C. Greub, T. Hurth and D. Wyler, \plb{380}{96}{385}; \prd{54}{96}{3350}.

\bibitem{misiak}
K.G. Chetyrkin, M. Misiak and M. M\"{u}nz, in preparation; talk given by M. 
Misiak at ICHEP96, Warsaw, Poland, July 1996.

\bibitem{hurth}
C. Greub and T. Hurth, SLAC preprint SLAC-PUB-7267, August 1996, 
hep-ph/9608449.

\bibitem{bertol}
S. Bertolini, F. Borzumati, A. Masiero and G. Ridolfi, \npb{353}{91}{591}.

\bibitem{barb}
R. Barbieri and G.F. Giudice, \plb{309}{93}{86};\\
N. Oshimo, \npb{404}{93}{20};\\
R. Garisto and J.N. Ng, \plb{315}{93}{372};\\
M.A. Diaz, \plb{304}{93}{278};\\
Y. Okada, \plb{315}{93}{119};\\
F. Borzumati, \zpc{63}{94}{291};\\
P. Nath and R. Arnowitt, \plb{336}{94}{395};\\
S. Bertolini and F. Vissani, \zpc{67}{95}{513};\\
J. Lopez et al., \prd{51}{95}{147}.

\bibitem{vissani}
S. Bertolini and F. Vissani, in ref.\cite{barb}.

\bibitem{cleo2}
R. Balest et al. (CLEO collab.), CLEO-CONF 94-4 (1994).

\bibitem{cdf}
C. Anway-Wiese et al. (CDF collab), Fermilab-Conf-95/201-E (1995).

\bibitem{ligeti}
Z. Ligeti and M.B. Wise, \prd{53}{96}{4937}.

\bibitem{cho}
P. Cho, M. Misiak and D. Wyler, \prd{54}{96}{3329}.

\bibitem{ali}
A. Ali, G.F. Giudice and T. Mannel, \zpc{67}{95}{417}.

\bibitem{kurimoto}
T. Kurimoto, \prd{39}{89}{3447}.

\bibitem{brignole}
G.C. Branco, G.C. Cho, Y. Kizukuri and N. Oshimo, \plb{337}{94}{316}; 
\npb{44}{95}{483};\\
A. Brignole, F. Feruglio and F. Zwirner, CERN preprint CERN-TH-95-340, 
hep-ph/9601293.

\bibitem{strumia}
R. Barbieri, L.J. Hall and A. Strumia, \npb{449}{95}{437};\\
N. Arkani-Hamed, H.-C. Cheng and L.J. Hall, \prd{53}{96}{413}.

\bibitem{kagan}
A.L. Kagan, \prd{51}{95}{6196}.

\bibitem{ciu}
M. Ciuchini, E. Gabrielli and G.F. Giudice, CERN preprint CERN-TH-96-073, 
April 1996, hep-ph/9604438.

\end{thebibliography}
\end{document}